\documentclass[12pt]{article}

\usepackage{cite,hyperref,amsfonts,amsmath,amssymb,bm,graphicx,a4wide}

\numberwithin{equation}{section}

\newcommand{\lyxdot}{.}

\begin{document}

\title{\textbf{Chiral magnetic effect in the presence of \\ an external axial-vector
field}}

\author{Maxim Dvornikov\thanks{maxdvo@izmiran.ru}
\\
\small{\ Pushkov Institute of Terrestrial Magnetism, Ionosphere} \\
\small{and Radiowave Propagation (IZMIRAN),} \\
\small{108840 Moscow, Troitsk, Russia;} \\
\small{\ Physics Faculty, National Research Tomsk State University,} \\
\small{36 Lenin Avenue, 634050 Tomsk, Russia}}

\date{}

\maketitle

%
%
%
%
%
%
%
%

\begin{abstract}
We study the excitation of the electric current of chiral fermions
along the external magnetic field, known as the chiral magnetic effect,
in the presence of the background axial-vector field. The calculation
of the current is based on the exact solution of the Dirac equation
for these fermions accounting for the external fields. First, this
solution was obtained for massive particles and, then, we consider
the chiral limit, which is used in the anomalous current computation.
We obtain that, in this situation, the anomalous current does not
contain the direct contribution of the axial-vector field. This result
is compared with findings of other authors.
\end{abstract}

\maketitle

\section{Introduction}

The evolution of chiral charged particles in external fields reveals
multiple quantum phenomena. First, we mention the Adler-Bell-Jackiw
anomaly~\cite{PesSch95}, which consists in the nonconservation
of the axial current in the presence of an external electromagnetic
field. This anomaly was shown in Ref.~\cite{NieNin81} to be closely
related to the excitation of the electric current of massless fermions $\mathbf{J}_{\mathrm{CME}}=\alpha_{\mathrm{em}}(\mu_{\mathrm{R}}-\mu_{\mathrm{L}})\mathbf{B}/\pi$ along the external magnetic field $\mathbf{B}$. Here $\alpha_{\mathrm{em}} \approx 1/137$
is the fine structure constant and $\mu_{\mathrm{R,L}}$ are the chemical
potentials of right and left chiral fermions. This phenomenon was named the chiral magnetic effect (the CME) later in Ref.~\cite{FukKhaWar08}. One can also mention
the chiral vortical effect (the CVE), described in Ref.~\cite{Vil79}, which is the generation
of the anomalous current in a rotating matter. There are active searches
for manifestations of the CME in astrophysics and cosmology~\cite{Sig17},
as well as in accelerator physics~\cite{Koc17}.

There is an open question on the influence of the external axial-vector
field $V_{5}^{\mu}$ on the magnitude of the anomalous current in
the CME. If a homogeneous and isotropic $V_{5}^{\mu}$ is present,
the Lagrangian of the interaction of the fermion field $\psi$ with
$V_{5}^{\mu}$ can be represented as $\mathcal{L}_{5}\sim\bar{\psi}\gamma^{\mu}\gamma^{5}\psi V_{5\mu}\to\psi^{\dagger}\gamma^{5}\psi V_{5}^0$,
which shows that the chiral imbalance $\mu_{5}=(\mu_{\mathrm{R}}-\mu_{\mathrm{L}})/2$
could be shifted by $V_{5}^0$. Here $\gamma^{\mu}=(\gamma^{0},\bm{\gamma})$
and $\gamma^{5}=\mathrm{i}\gamma^{0}\gamma^{1}\gamma^{2}\gamma^{3}$
are the Dirac matrices. Thus the CME could contain the contribution
of $V_{5}^0\equiv V_{5}$. This idea was recently implemented in
Ref.~\cite{BubGubZhu17}. However the regularization used in Ref.~\cite{BubGubZhu17}
to compute the divergent integrals was ambiguous.

Another possibility for $V_{5}^{\mu}$ to affect the CME is the consideration
of the polarization operator of a photon in the fermion plasma under
the influence of $V_{5}^{\mu}$. In this case, the polarization operator
could acquire the additional term $\Pi_{\mu\nu}\sim\mathrm{i}\varepsilon_{\mu\nu\lambda\rho}V_{5}^{\lambda}k^{\rho}$,
where $k^{\rho}$ is the photon momentum and $\varepsilon_{\mu\nu\lambda\rho}$
is the antisymmetric tensor. The appearance of this term is equivalent
to the excitation of the current $\mathbf{J}\sim V_{5}\mathbf{B}$,
which would be a direct contribution of $V_{5}^{\mu}$ to the CME.

The calculation of the antisymmetric contribution to $\Pi_{\mu\nu}$ in the QED plasma of chiral fermions was made in Ref.~\cite{AkaYam13}, where the CME was reproduced. Then, analogous calculations in the presence of an external axial-vector field, e.g., the electroweak interaction with background matter, were carried out in Refs.~\cite{BoyRucSha12,DvoSem14}. The calculations in Refs.~\cite{BoyRucSha12,DvoSem14} demonstrate that one has the nonzero contribution to the polarization tensor $\Pi_{ij} \sim \mathrm{i}\varepsilon_{ijn}k^{n} V_5$, showing that the CME can be influenced by the external axial-vector field $V_{5}^{\mu}$. However, a model with $\mathcal{L}_{5}\neq0$ is nonrenormalizable
and the result of the calculation of $\Pi_{\mu\nu}$ was shown in
Ref.~\cite{JacKos99} to depend on the regularization scheme applied.

Using the perturbative loop expansion, the CVE was found in Ref.~\cite{GolSon15} to receive no radiative corrections from Yukawa-type interactions. This result can be extended to the CME. However, the arguments against the corrections to the CVE are not applicable in the case of dynamical gauge fields, which can be present in a realistic system.
We can also mention that lattice calculations, performed in Ref.~\cite{BuiPuhVal},
show that the CME can get a contribution from an interfermion interaction.

The aim of this work is to find out whether there is an influence
of the axial-vector field on the CME. We consider a particular
example of this axial-vector field in the form of the electroweak
interaction with background matter. We start in Sec.~\ref{sec:MOTIV}
with the description of the motivation for this study. Then, in Sec.~\ref{sec:CANCCURR},
we calculate the anomalous current along the external magnetic field
basing on the exact solution of the Dirac equation in the external
fields. Our results are discussed in Sec.~\ref{sec:DISC}.

\section{Motivation\label{sec:MOTIV}}

In this section, we compare the results of different methods for the
calculation of $\mathbf{J}_{\mathrm{CME}}\parallel\mathbf{B}$ in
the presence of $V_{5}^{\mu}$, which is taken in the form of the electroweak interaction with
background matter.

The method of the relativistic quantum mechanics was used for the
first time to describe the CME in Ref.~\cite{Vil80}. The idea of
this method is the following. First, one obtains the exact solution
of the Dirac equation for a massless particle in an external magnetic
field. Then the electric current is computed as
\begin{equation}\label{eq:currabstr}
  \mathbf{J}_{\chi}=q
  \left\langle
    \bar{\psi}_{\chi}\bm{\gamma}\psi_{\chi}
  \right\rangle,
\end{equation}
where $\psi_{\chi}$ is the wave function, obtained in the solution
of the Dirac equation and $q$ is the particle charge. The averaging
$\left\langle \dots\right\rangle $ in Eq.~(\ref{eq:currabstr})
is made over the statistical ensemble. The contribution of any chirality
$\chi=\mathrm{L,R}$ is accounted for in Eq.~(\ref{eq:currabstr}).
This method allows one to take into account the contribution of the
external field nonperturbatively. Nevertheless, it is restricted to
the constant and homogeneous magnetic field.

The application of the relativistic quantum mechanics method for the
study of the CME in the presence of the electroweak parity violating
interaction was made in Refs.~\cite{DvoSem15a,DvoSem15b}. Let us
consider a massless electron, electroweakly interacting with nonmoving
and unpolarized background matter under the influence of an external
magnetic field along the $z$ axis, $\mathbf{B}=B\mathbf{e}_{z}$.
The Lagrangian for such an electron, described by the bispinor $\psi_{e}$,
has the form,
\begin{equation}\label{eq:Largchir}
  \mathcal{L}=\bar{\psi}_{e}
  \left[
    \gamma_{\mu}(\mathrm{i}\partial^{\mu}+eA^{\mu})-
    \gamma_{0}(V_{\mathrm{L}}P_{\mathrm{L}}+V_{\mathrm{R}}P_{\mathrm{R}})
  \right]
  \psi_{e},
\end{equation}
where $A^{\mu}=(0,0,Bx,0)$ is the vector potential corresponding
to the constant and homogeneous magnetic field, $e>0$ is the elementary
charge, $P_{\mathrm{L,R}}=(1\mp\gamma^{5})/2$ are the chiral projectors,
and $V_{\mathrm{L,R}}$ are the effective potentials of the electroweak
interaction of the electron chiral projections with background matter.
The explicit form of $V_{\mathrm{L,R}}$ is given in Ref.~\cite{DvoSem15a}
for the case of background matter consisting of neutrons and protons.

The energy spectra of the chiral projections of the electron were
found in Refs.~\cite{DvoSem15a,DvoSem15b} as
\begin{equation}\label{eq:spectrumwrong}
  E_{\mathrm{L,R}}=V_{\mathrm{L,R}}+\mathcal{E}_{0},
  \quad
  \mathcal{E}_{0}=\sqrt{p_{z}^{2}+2eBn},
\end{equation}
where $p_{z}$ is the longitudinal momentum along the magnetic field,
and $n=0,1,\dots$ is the discrete quantum number. If $n=0$ in Eq.~(\ref{eq:spectrumwrong}),
we have obtained in Refs.~\cite{DvoSem15a,DvoSem15b} for left electrons,
\begin{equation}\label{eq:ELwrong}
  E_{\mathrm{L}}^{(n=0)}=V_{\mathrm{L}}+p_{z},\quad0<p_{z}<+\infty,
\end{equation}
and for right particles,
\begin{equation}\label{eq:ERwrong}
  E_{\mathrm{R}}^{(n=0)}=V_{\mathrm{R}}-p_{z},\quad-\infty<p_{z}<0.
\end{equation}
In Eqs.~(\ref{eq:ELwrong}) and~(\ref{eq:ERwrong}) we assumed,
in analogy with the situation when $V_{\mathrm{L,R}}=0$, that
left and right electrons move in a certain direction along the magnetic
field.

Using the spectra in Eqs.~(\ref{eq:spectrumwrong})-(\ref{eq:ERwrong})
and the wave functions, found in Refs.~\cite{DvoSem15a,DvoSem15b},
as well as applying the general expression for the current in Eq.~(\ref{eq:currabstr}),
one finds that only the lowest energy level with $n=0$ contributes to
the current, giving one its nonzero component along the magnetic field
as
\begin{equation}\label{eq:Jwrong}
  \mathbf{J}=\frac{2\alpha_{\mathrm{em}}}{\pi}(\mu_{5}+V_{5})\mathbf{B},
\end{equation}
where $\alpha_{\mathrm{em}}=e^{2}/4\pi$ is the fine structure constant
and $V_{5}=(V_{\mathrm{L}}-V_{\mathrm{R}})/2$ is the contribution
of the electroweak interaction.

The result in Eq.~(\ref{eq:Jwrong}) was criticized in Ref.~\cite{KapRedSen17},
where it was found that
\begin{equation}\label{eq:Jcorr}
  \mathbf{J}=\frac{2\alpha_{\mathrm{em}}}{\pi}\mu_{5}\mathbf{B} \equiv
  \mathbf{J}_{\mathrm{CME}},
\end{equation}
even in the presence of the background electroweak matter, i.e. when
$V_{\mathrm{L,R}}\neq0$. The authors of Ref.~\cite{KapRedSen17}
used the alternative derivation of the CME based on the energy balance
in the motion of a massless charged particle in parallel electric
and magnetic fields, previously proposed in Ref.~\cite{NieNin81}. %
Note that, even if one uses the method of Ref.~\cite{NieNin81}
to derive the CME in the presence the electroweakly interacting matter
and accounts for the dispersion relation in Eqs.~(\ref{eq:ELwrong})
and~(\ref{eq:ERwrong}), the expression for the current in Eq.~(\ref{eq:Jwrong})
can be reproduced~\cite{DvoSem18}.

The expression for the current in Eq.~(\ref{eq:Jcorr}) was also
derived in Ref.~\cite{SadIsa11}, where the CME in the presence of
the axial-vector field was studied using the chiral hydrodynamics
approach, which was developed earlier in Ref.~\cite{SonSur09}. No
explicit contribution of the electroweak interaction to the current,
like in Eq.~(\ref{eq:Jwrong}), was found in Ref.~\cite{SadIsa11}.

The apparent contradiction between the relativistic quantum mechanics
method in Refs.~\cite{DvoSem15a,DvoSem15b} and other approaches~\cite{KapRedSen17,SadIsa11}
for the description of the CME in the presence of the axial-vector
external field requires a special analysis.

\section{Anomalous current in the presence of the electroweak interaction
with matter\label{sec:CANCCURR}}

To analyze the contribution of the parity violating interaction to
the CME, using the relativistic quantum mechanics approach, we
start with the consideration of massive particles interacting with
the axial-vector and magnetic fields and then discuss the chiral limit.
As in Sec.~\ref{sec:MOTIV}, we consider the electroweak interaction
of an electron with background matter. Since the relativistic quantum
mechanics method deals with an exact solution of the Dirac equation
in an external field, the corresponding solution should be utilized.
For the first time the Dirac equation for a massive electron, electroweakly
interacting with background matter under the influence of an external
magnetic field, was solved in Ref.~\cite{BalPopStu11}. Then this
solution was used in Ref.~\cite{Dvo16a} to compute the induced current
along the magnetic field.

Thus, instead of the Lagrangian in Eq.~(\ref{eq:Largchir}), we 
discuss the following Lagrangian:
\begin{equation}\label{eq:Largmass}
  \mathcal{L}=\bar{\psi}_{e}
  \left[
    \gamma_{\mu}(\mathrm{i}\partial^{\mu}+eA^{\mu})-m-
    \gamma_{0}(V_{\mathrm{L}}P_{\mathrm{L}}+V_{\mathrm{R}}P_{\mathrm{R}})
  \right]\psi_{e},
\end{equation}
where $m$ is the electron mass. The remaining parameters have the
same meaning as in Sec.~\ref{sec:MOTIV}.

Let us look for the solution of the Dirac equation, which results
from Eq.~(\ref{eq:Largmass}), in the form,
\begin{equation}\label{eq:psiel}
  \psi_{e}=\exp
  \left(
    -\mathrm{i}Et+\mathrm{i}p_{y}y+\mathrm{i}p_{z}z
  \right)
  \psi_{x},
\end{equation}
where $\psi_{x}=\psi(x)$ is the bispinor which depends on $x$ and
$p_{y,z}$ are the momentum projections along the $y$ and $z$ axes.
We choose the chiral representation of the Dirac matrices~\cite{ItzZub80},
\begin{equation}\label{eq:chirrep}
  \gamma^{\mu}=
  \begin{pmatrix}
    0 & -\sigma^{\mu}\\
    -\bar{\sigma}^{\mu} & 0\ 
  \end{pmatrix},
  \quad\sigma^{\mu}=(\sigma_{0},-\bm{\sigma}),
  \quad
  \bar{\sigma}^{\mu}=(\sigma_{0},\bm{\sigma}),
\end{equation}
where $\sigma_{0}$ is the unit $2\times2$ matrix and $\bm{\sigma}$
are the Pauli matrices. Using Eq.~(\ref{eq:chirrep}), we can represent
$\psi_{x}$ in the form~\cite{Dvo16a}
\begin{equation}\label{eq:psix}
  \psi_{x}^{\mathrm{T}}=
  \left(
    C_{1}u_{n-1},\mathrm{i}C_{2}u_{n},C_{3}u_{n-1},\mathrm{i}C_{4}u_{n}
  \right),
\end{equation}
where $C_{i}$, $i=1,\dots,4,$ are the spin coefficients,
\begin{equation}\label{eq:Hermfun}
  u_{n}(\eta)=
  \left(
    \frac{eB}{\pi}
  \right)^{1/4}
  \exp
  \left(
    -\frac{\eta^{2}}{2}
  \right)\frac{H_{n}(\eta)}{\sqrt{2^{n}n!}},
  \quad
  n=0,1,\dotsc,
\end{equation}
are the Hermite functions, $H_{n}(\eta)$ are the Hermite polynomials,
and $\eta=\sqrt{eB}x+p_{y}/\sqrt{eB}$.

The energy spectrum for $n>0$ reads~\cite{BalPopStu11,Dvo16a}
\begin{equation}\label{eq:En>0}
  E=\bar{V}+\lambda\mathcal{E},
  \quad
  \mathcal{E}=\sqrt{(\mathcal{E}_{0}+sV_{5})^{2}+m^{2}},
\end{equation}
where $s=\pm1$ is the discrete quantum number dealing with the spin
operator~\cite{BalPopStu11}, $\mathcal{E}_{0}$ is defined in Eq.~(\ref{eq:spectrumwrong}),
$\bar{V}=(V_{\mathrm{L}}+V_{\mathrm{R}})/2$, and $\lambda=\pm1$
is the sign of the energy; i.e. the electron energy reads $E_{e}=E(\lambda=+1)=\mathcal{E}+\bar{V}$,
and the positron energy has the form $E_{\bar{e}}=-E(\lambda=-1)=\mathcal{E}-\bar{V}$.
For $n=0$, one has~\cite{Dvo16a}
\begin{equation}\label{eq:Elowest}
  E=\bar{V}+\lambda\mathcal{E},
  \quad
  \mathcal{E}=\sqrt{(p_{z}+V_{5})^{2}+m^{2}}.
\end{equation}
Note that, at $n=0$, there is only one spin state of the electron.

The spin coefficients obey the system~\cite{Dvo16a},
\begin{align}\label{eq:Cisys}
  \left(
    \mathcal{E}\mp p_{z}\pm V_{5}
  \right)
  C_{1,3}\mp\sqrt{2eBn}C_{2,4}+mC_{3,1} & =0,  
  \nonumber
  \\
  \left(
    \mathcal{E}\pm p_{z}\pm V_{5}
  \right)
  C_{2,4}\mp\sqrt{2eBn}C_{1,3}+mC_{4,2} & =0,
\end{align}
where we consider the particle (electron) degrees of freedom, $\lambda=1$.
Since we are mainly interested in the dynamics of electrons at the
lowest energy level, we should set $n=0$ in Eq.~(\ref{eq:Cisys}).
It results from Eq.~(\ref{eq:psix}) that, in this situation, $C_{1}=C_{3}=0$
to avoid the appearance of Hermite functions with negative indices.

If, besides setting $n=0$ in Eq.~(\ref{eq:Cisys}), we approach
to the limit $m\to0$ there, one gets
\begin{align}
  \left(
    \mathcal{E}+p_{z}+V_{5}
  \right)
  C_{2}= & 0,
  \quad
  \text{or}
  \quad
  \begin{cases}
    \mathcal{E}=-p_{z}-V_{5},
    \\
    C_{2}\neq0,
    \quad
    \text{and}
    \quad
    C_{4}=0,
  \end{cases}
  \label{eq:Rraw}
  \\
  \left(
    \mathcal{E}-p_{z}-V_{5}
  \right)
  C_{4}= & 0,
  \quad
  \text{or}
  \quad
  \begin{cases}
    \mathcal{E}=p_{z}+V_{5},
    \\
    C_{4}\neq0,
    \quad
    \text{and}
    \quad
    C_{2}=0.
  \end{cases}
  \label{eq:Lraw}
\end{align}
We can see that Eq.~(\ref{eq:Rraw}) corresponds to a right electron
and Eq.~(\ref{eq:Lraw}) to a left one.

The energy spectrum in Eq.~(\ref{eq:Elowest}) in the limit $m\to0$
reads
\begin{equation}\label{eq:chirspecgen}
  \mathcal{E}=|p_{z}+V_{5}|.
\end{equation}
Comparing Eq.~(\ref{eq:chirspecgen}) with Eqs.~(\ref{eq:Rraw})
and~(\ref{eq:Lraw}), we obtain that for a right electron
\begin{equation}\label{eq:Rrange}
  |p_{z}+V_{5}|=-p_{z}-V_{5},\quad\text{or}\quad p_{z}<-V_{5},
\end{equation}
and
\begin{equation}\label{eq:Lrange}
  |p_{z}+V_{5}|=p_{z}+V_{5},\quad\text{or}\quad p_{z}>-V_{5},
\end{equation}
for a left particle.

Therefore the total energy of a left electron at the lowest energy
level has the form
\begin{equation}\label{eq:ELcorrect}
  E_{e\mathrm{L}}^{(n=0)}=V_{\mathrm{L}}+p_{z},
  \quad
  -V_{5}<p_{z}<+\infty,
\end{equation}
and
\begin{equation}\label{eq:ERcorrect}
  E_{e\mathrm{R}}^{(n=0)}=V_{\mathrm{R}}-p_{z},
  \quad
  -\infty<p_{z}<-V_{5},
\end{equation}
of a right particle. Comparing Eqs.~(\ref{eq:ELcorrect}) and~(\ref{eq:ERcorrect})
with Eqs.~(\ref{eq:ELwrong}) and~(\ref{eq:ERwrong}), we can see
that the form of the spectrum at $n=0$, obtained here, formally coincides
with that used in Refs.~\cite{DvoSem15a,DvoSem15b}. However, the
range of the $p_{z}$ variation is different.

To complete the solution of the Dirac equation at $n=0$ and $m\to0$
we should fix the remaining spin coefficients. One gets that
\begin{equation}\label{eq:C24}
  C_{2}^{(\mathrm{R})}=C_{4}^{(\mathrm{L})}=\frac{1}{2\pi},
  \quad
  C_{2}^{(\mathrm{L})}=C_{4}^{(\mathrm{R})}=0,
\end{equation}
which results from the normalization condition
\begin{equation}\label{eq:psinorm}
  \int\mathrm{d}^{3}x\psi_{p_{y}p_{z}n}^{\dagger}\psi_{p'_{y}p'_{z}n'} =
  \delta
  \left(
    p_{y}-p'_{y}
  \right)
  \delta
  \left(
    p_{z}-p'_{z}
  \right)
  \delta_{nn'},
\end{equation}
of the total wave function.

The wave function of a positron can be obtained from Eqs.~(\ref{eq:psiel})
and~(\ref{eq:psix}) by applying the charge conjugation $\psi_{\bar{e}}=\mathrm{i}\gamma^{2}\psi_{e}^{*}$
and setting $\lambda=-1$ in Eq.~(\ref{eq:En>0}). Finally one has
\begin{equation}\label{eq:psipos}
  \psi_{\bar{e}}^{\mathrm{T}}= 
  \exp(-\mathrm{i}E_{\bar{e}}t-\mathrm{i}p_{y}y-\mathrm{i}p_{z}z)
  \times
  \left(
    -\mathrm{i}C_{4}u_{n},-C_{3}u_{n-1},\mathrm{i}C_{2}u_{n},C_{1}u_{n-1}
  \right),
\end{equation}
where the coefficients $C_{i}$ obey the system in Eq.~(\ref{eq:Cisys}).

If $n=0$, we obtain on the basis of Eqs.~(\ref{eq:psipos}) and~(\ref{eq:Elowest})
that
\begin{equation}\label{eq:psiposR}
  \psi_{\bar{e}\mathrm{R}}^{(n=0)}=
  \exp(-\mathrm{i}E_{\bar{e}\mathrm{R}}t-\mathrm{i}p_{y}y-\mathrm{i}p_{z}z)
  \times
  \frac{\mathrm{i}u_{0}}{2\pi}
  \left(
    -1,0,0,0
  \right)^{\mathrm{T}},
\end{equation}
where
\begin{equation}\label{eq:EposR}
  E_{\bar{e}\mathrm{R}}^{(n=0)}=p_{z}-V_{\mathrm{R}},
  \quad
  -V_{5}<p_{z}<+\infty,
\end{equation}
is the energy of right positrons at the lowest energy level. For left
positrons one has
\begin{equation}\label{eq:psiposL}
  \psi_{\bar{e}\mathrm{L}}^{(n=0)}=
  \exp(-\mathrm{i}E_{\bar{e}\mathrm{L}}t-\mathrm{i}p_{y}y-\mathrm{i}p_{z}z)
  \times
  \frac{\mathrm{i}u_{0}}{2\pi}
  \left(
    0,0,1,0
  \right)^{\mathrm{T}},
\end{equation}
where
\begin{equation}\label{eq:EposL}
  E_{\bar{e}\mathrm{L}}^{(n=0)}=-p_{z}-V_{\mathrm{L}},
  \quad
  -\infty<p_{z}<-V_{5},
\end{equation}
is the energy of left positrons at the lowest energy level. The positron
wave functions in Eqs.~(\ref{eq:psiposR}) and~(\ref{eq:psiposL})
satisfy the normalization condition in Eq.~(\ref{eq:psinorm}).

\begin{figure}
  \centering
  \includegraphics[scale=0.3]{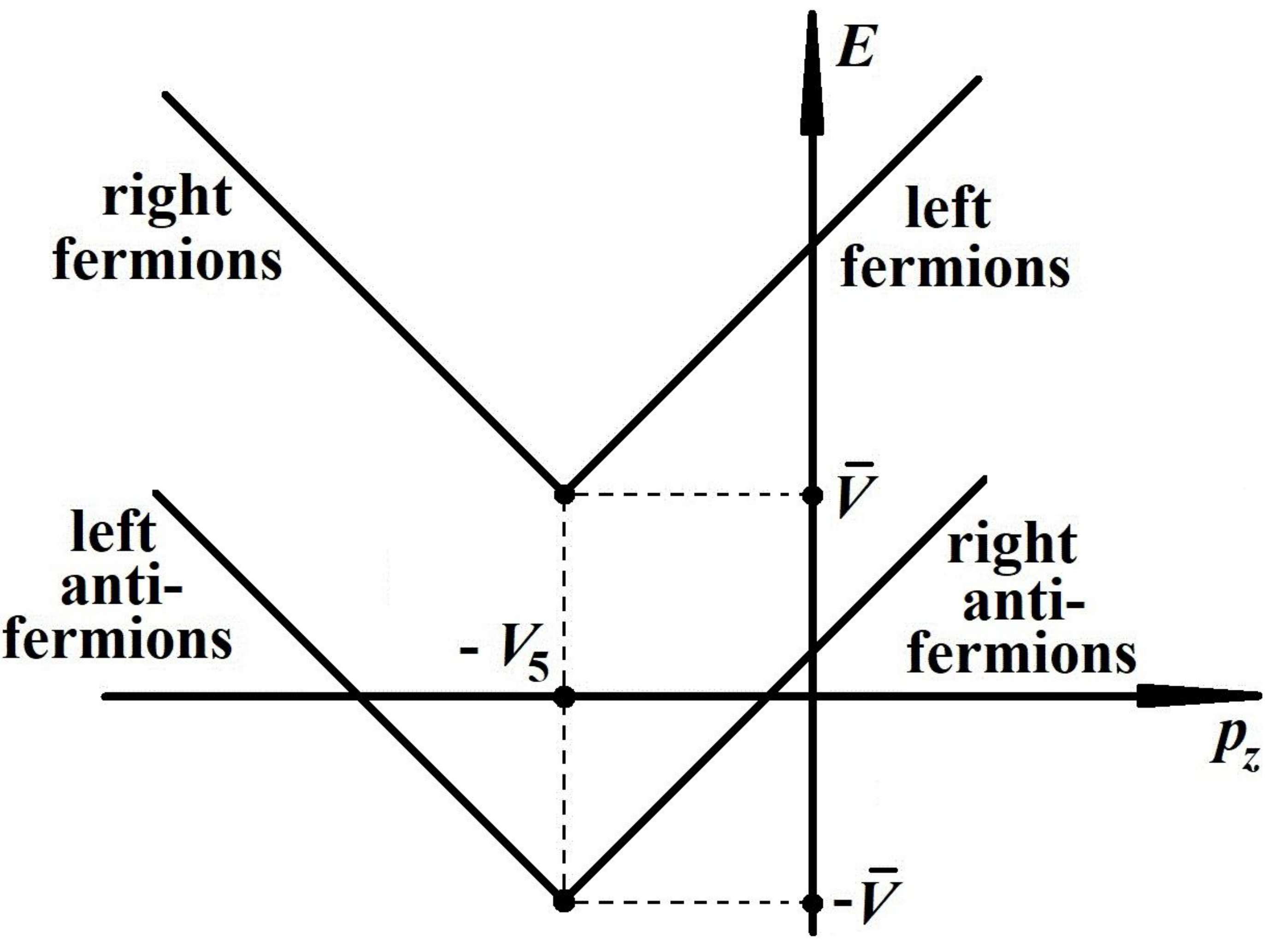}
  \caption{The energy spectrum of massless left and right electrons/positrons
  at the lowest energy level with $n=0$ electroweakly interacting with
  background matter. This plot corresponds to Eqs.~(\ref{eq:ELcorrect}),
  (\ref{eq:ERcorrect}), (\ref{eq:EposR}), and~(\ref{eq:EposL}).\label{fig:enspec}}
\end{figure}

The energy spectrum for electrons and positrons at the lowest energy
level is shown in Fig.~\ref{fig:enspec}. One can see that there
is no gap between the dispersion relations of left and right electrons/positrons,
predicted in Refs.~\cite{DvoSem15a,DvoSem15b,DvoSem18}; i.e. the
assumption that $E_{\mathrm{Lmin}}^{(n=0)}\neq E_{\mathrm{Rmin}}^{(n=0)}$
at $p_{z}=0$ [see Eqs.~(\ref{eq:ELwrong}) and~(\ref{eq:ERwrong})]
is incorrect. In the presence of the electroweak matter, the spectrum
of massless electrons/positrons with $n=0$ is parallel transported
to the point $(p_{z}=-V_{5},E=\bar{V})$ for electrons and to $(p_{z}=-V_{5},E=-\bar{V})$
for positrons from the point $(p_{z}=0,E=0)$ corresponding to the
vacuum case.

According to Eq.~(\ref{eq:currabstr}), the contributions of left
and right electrons at the lowest energy level to the current are
\begin{equation}\label{eq:JLRgen}
  \mathbf{J}_{e\mathrm{L,R}}^{(n=0)}=
  -e\int\mathrm{d}p_{y}\mathrm{d}p_{z}
  \bar{\psi}_{e  \mathrm{L,R}}\bm{\gamma}\psi_{e\mathrm{L,R}}
  f(E_{e\mathrm{L,R}}^{(n=0)}-\mu_{\mathrm{L,R}}),
\end{equation}
where $f(E)=\left[\exp(\beta E)+1\right]^{-1}$ is the Fermi-Dirac
distribution function, $\beta=1/T$ is the reciprocal temperature, and 
$\mu_{\mathrm{L,R}}$ are the chemical potentials of left and right
particles. First we notice that the components of the current, transverse
with respect to $\mathbf{B}$, are vanishing. Performing the integration
over $-\infty<p_{y}<+\infty$ and accounting for Eqs.~(\ref{eq:ELcorrect})-(\ref{eq:C24}),
on the basis of Eq.~(\ref{eq:JLRgen}) we obtain the expression for
the total current of electrons $\mathbf{J}_{e}^{(n=0)}=\mathbf{J}_{\mathrm{L}}^{(n=0)}+\mathbf{J}_{\mathrm{R}}^{(n=0)}$
at $n=0$, 
\begin{align}\label{eq:Je}
  \mathbf{J}_{e}^{(n=0)}= &
  \frac{e^{2}\mathbf{B}}{(2\pi)^{2}}
  \left[
    \int_{-\infty}^{-V_{5}}\mathrm{d}p_{z}f(-p_{z}+V_{\mathrm{R}}-\mu_{\mathrm{R}})- 
    \int_{-V_{5}}^{+\infty}\mathrm{d}p_{z}f(p_{z}+V_{\mathrm{L}}-\mu_{\mathrm{L}})
  \right]
  \nonumber
  \\
  & =
  \frac{e^{2}\mathbf{B}}{(2\pi)^{2}}\int_{0}^{+\infty}\mathrm{d}p
  \left[
    f(p+\bar{V}-\mu_{\mathrm{R}})-f(p+\bar{V}-\mu_{\mathrm{L}})
  \right].
\end{align}
Analogously to Refs.~\cite{DvoSem15a,DvoSem15b} one can show that
higher energy levels with $n>0$ do not contribute to the current.
Thus we omit the superscript in Eq.~(\ref{eq:Je}) for brevity.

The positron contribution to the current $\mathbf{J}_{\bar{e}}$ can
be obtained analogously to Eq.~(\ref{eq:JLRgen}) as
\begin{equation}\label{eq:JLRpos}
  \mathbf{J}_{\bar{e}\mathrm{L,R}}^{(n=0)}=
  e\int\mathrm{d}p_{y}\mathrm{d}p_{z}
  \bar{\psi}_{\bar{e}\mathrm{L,R}}\bm{\gamma}\psi_{\bar{e}\mathrm{L,R}}
  f(E_{\bar{e}\mathrm{L,R}}^{(n=0)}+\mu_{\mathrm{L,R}}).
\end{equation}
Using Eqs.~(\ref{eq:psiposR})-(\ref{eq:EposL}), we derive, on the
basis of Eq.~(\ref{eq:JLRpos}),
\begin{align}\label{eq:Jpos}
  \mathbf{J}_{\bar{e}}^{(n=0)}= & \frac{e^{2}\mathbf{B}}{(2\pi)^{2}}
  \left[
    \int_{-\infty}^{-V_{5}}\mathrm{d}p_{z}f(-p_{z}-V_{\mathrm{L}}+\mu_{\mathrm{L}})
    -\int_{-V_{5}}^{+\infty}\mathrm{d}p_{z}f(p_{z}-V_{\mathrm{R}}+\mu_{\mathrm{R}})  
  \right]
  \nonumber
  \\
  & =
  \frac{e^{2}\mathbf{B}}{(2\pi)^{2}}\int_{0}^{+\infty}\mathrm{d}p
  \left[
    f(p-\bar{V}+\mu_{\mathrm{L}})-f(p-\bar{V}+\mu_{\mathrm{R}})
  \right],
\end{align}
for the total contribution to the current from positrons at the lowest energy level.


Using Eqs.~(\ref{eq:Je}) and~(\ref{eq:Jpos}), we obtain that the
total current $\mathbf{J}=\mathbf{J}_{e}+\mathbf{J}_{\bar{e}}$ reads
\begin{align}\label{eq:Jfinal}
  \mathbf{J}= & \frac{e^{2}\mathbf{B}}{(2\pi)^{2}}\int_{0}^{+\infty}\mathrm{d}p
  \left[
    f(p+\bar{V}-\mu_{\mathrm{R}})-f(p-\bar{V}+\mu_{\mathrm{R}})-
    f(p+\bar{V}-\mu_{\mathrm{L}})+f(p-\bar{V}+\mu_{\mathrm{L}})
  \right]
  \nonumber
  \\
  & =
  \frac{2\alpha_{\mathrm{em}}}{\pi}\mu_{5}\mathbf{B},
\end{align}
which is in agreement with Eq.~(\ref{eq:Jcorr}).

\section{Discussion\label{sec:DISC}}

In the present work, we have elaborated the improved derivation of
the anomalous current of massless charged fermions, interacting with
an axial-vector field under the influence of the external magnetic
field, induced along this magnetic field. We have chosen a particular
example of the axial-vector field as the electroweak interaction of
an electron with nonmoving and unpolarized background matter. Unlike
Refs.~\cite{KapRedSen17,SadIsa11}, here we have used the method
of the relativistic quantum mechanics, originally proposed in Ref.~\cite{Vil80}
to describe the CME.

Utilizing the exact solution of the Dirac equation, found in Refs.~\cite{BalPopStu11,Dvo16a},
we have shown in Sec.~\ref{sec:CANCCURR} that the axial-vector field
does not contribute to the current $\mathbf{J}\parallel\mathbf{B}$;
cf. Eq.~(\ref{eq:Jfinal}). The value of the current coincides with
the prediction of the CME in Eq.~(\ref{eq:Jcorr}) even in the case
when chiral fermions electroweakly interact with background matter,
confirming the findings of Refs.~\cite{KapRedSen17,SadIsa11}.

\begin{figure}
  \includegraphics[scale=0.3]{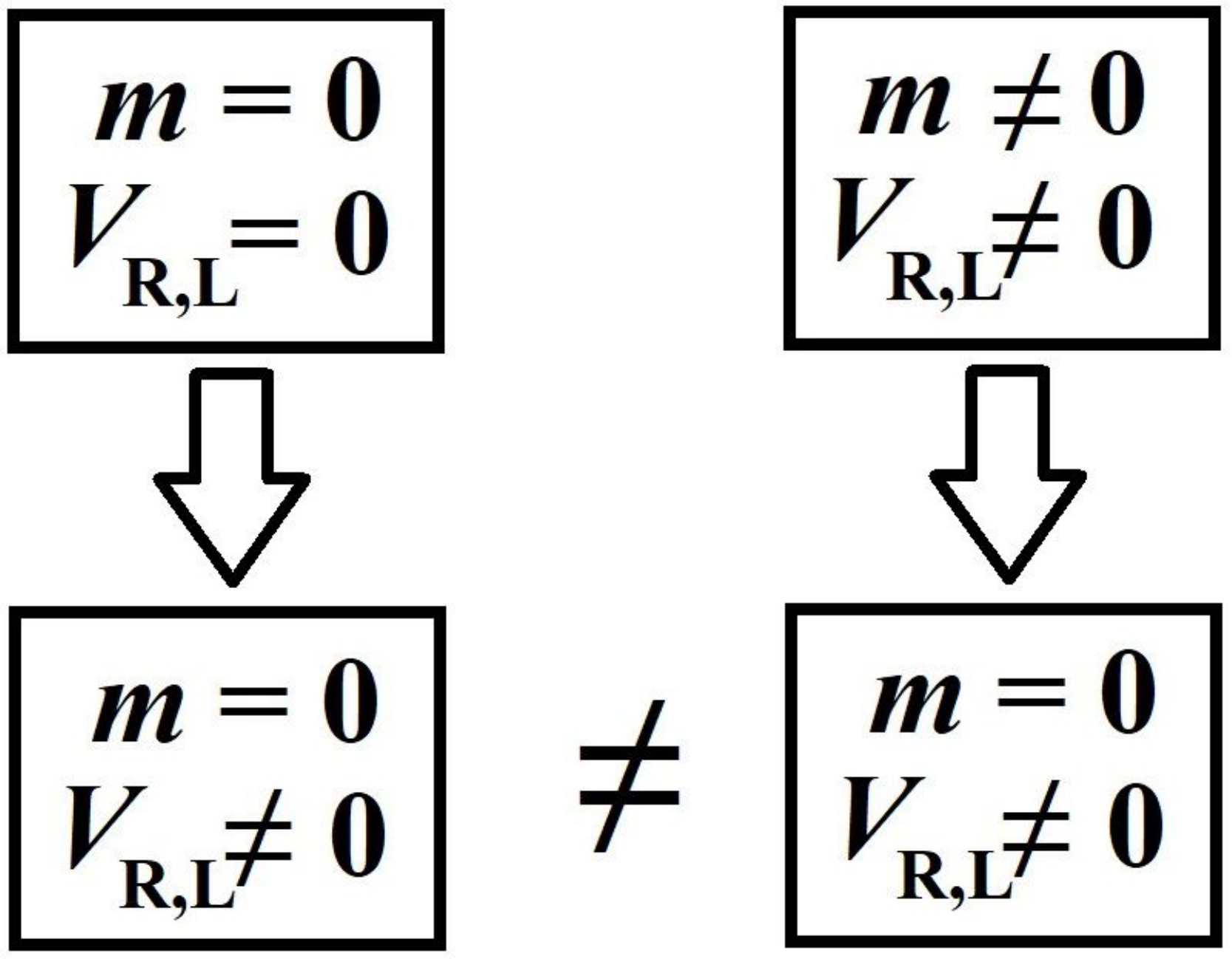}
  \centering
  \caption{Schematic diagram showing that the system with $m=0$ and
  $V_{\mathrm{R,L}}  \protect\neq0$,
  which is studied while considering the CME, can be prepared in two
  nonequivalent ways. One can start with a system of massless particles
  without the electroweak interaction (upper left box) and then turn
  on the electroweak interaction (lower left box). This scenario is
  implemented in Refs.~\cite{DvoSem15a,DvoSem15b,DvoSem18} and gives
  the current in Eq.~(\ref{eq:Jwrong}). However, this approach is
  not equivalent to that, where one, first, starts with massive particles
  with the electroweak interaction (upper right box) and then ``turns
  off'' the particle mass (lower right box). The latter approach results
  in the current in Eq.~(\ref{eq:Jcorr}).\label{fig:noncom}}
\end{figure}

To obtain this result in frames of the relativistic quantum mechanics
one has to consider the solution of the Dirac equation for a massive
electron in the external fields and then approach the limit $m\to0$.
If one sets $m=0$ in the Dirac equation from the very beginning,
i.e. if one considers the chiral Lagrangian in Eq.~(\ref{eq:Largchir}),
one obtains the current in Eq.~(\ref{eq:Jwrong}) as in Refs.~\cite{DvoSem15a,DvoSem15b},
which is inconsistent with the results of Refs.~\cite{KapRedSen17,SadIsa11}.
Thus we conclude that the system of chiral fermions, where the external
axial-vector field is present, can be prepared in two nonequivalent
ways. This fact is reflected in Fig.~\ref{fig:noncom}.

It is also interesting to notice that, at $n=0$, particles of a certain
chirality, say left-handed, are indirectly affected by the parameters
corresponding to the opposite (i.e., right-handed) chirality. It can
be seen in Eqs.~(\ref{eq:ELcorrect}) and~(\ref{eq:ERcorrect}).
Indeed, if one adiabatically changes $V_{\mathrm{L}}$, not only $E_{\mathrm{L}}^{(n=0)}$
but also $E_{\mathrm{R}}^{(n=0)}$ will be modified since the range
of the $p_{z}$ variation in $E_{\mathrm{R}}^{(n=0)}$ depends on
$V_{5}$. One would naively expect that left- and right-handed electrons behave
totally independently for purely massless particles.

The effect of the change of the particle momentum in the presence
of the electroweak interaction [see Eq.~(\ref{eq:chirspecgen})]
was known previously. In the case when a particle moves in a background
electroweak matter, both particle energy and its momentum acquire
the contributions $\sim G_{\mathrm{F}}$, where $G_{\mathrm{F}}$
is the Fermi constant. It results, e.g., in the appearance of the
ponderomotive force in the situation of the anisotropic matter with
inhomogeneous density~\cite{Sil99}.

We also mention that one does not need to involve the concept of
the Chern-Simons current~\cite{Lan16}, as suggested in Ref.~\cite{KapRedSen17},
to reconcile the results for the derivation of the CME in the presence
of the axial-vector field based on the relativistic quantum mechanics~\cite{Vil80}
and the energy balance arguments~\cite{NieNin81}. We can obtain
the coinciding results just by using the correct energy spectrum of
massless electrons at the lowest energy level; cf. Eqs.~(\ref{eq:ELcorrect})
and~(\ref{eq:ERcorrect}).

Since the electroweak interaction of chiral electrons with a homogeneous neutron matter does not contribute to the CME, one cannot expect the instability of the magnetic field and the amplification of the field in a neutron star (NS) to the magnetar strength, predicted in Refs.~\cite{DvoSem15a,DvoSem15b}. At the absence of $V_5$-contribution to the CME, one has the following qualitative behavior of the magnetic field in NS. If the initial chiral imbalance $\mu_5$ of ultrarelativistic electrons is present in the system, it is washed out very rapidly because of the helicity flip in electron collisions. The chiral imbalance does not recover since there is no $V_5$ driver in the $\mu_5$ evolution equation. Since $\mu_5 \to 0$, the densities of the magnetic energy and the magnetic helicity are not affected by the CME. Therefore the evolution of the magnetic field does not reveal an instability. Hence the magnetic field in NS can only experience a slow exponential decay because of the finite electric conductivity of NS matter. The direct numerical simulations confirm this magnetic field behavior.

Since the model of the magnetic field generation in magnetars driven by the electroweak interaction of electrons with background nucleons is likely to be invalid, other mechanisms for the explanation of magnetic fields in magnetars, different from that proposed in Refs.~\cite{DvoSem15a,DvoSem15b}, should be put forward. The classical magnetohydrodynamics (MHD) dynamo is unlikely to be sufficient for the generation of strong magnetic fields in magnetars~\cite{MerPonMel15}. Hence one should probably look for the solution of the magnetars problem in frames of the elementary particle physics, e.g., involving chiral phenomena.

In this respect, we can mention Ref.~\cite{Yam16a}, where the CVE is used to generate the fluid helicity of plasma under the influence of strong neutrino fluxes in a nascent NS. Then this fluid helicity can be converted to a helical magnetic field. To solve the problem of the magnetic field scale, it was suggested in Ref.~\cite{Yam16b} to implement the inverse cascade in chiral MHD in proto-NS matter. The neutrino driven creation of $\mu_5$ outside the neutrino-sphere in proto-NS was proposed in Ref.~\cite{SigLei16}. The generation of magnetic fields in a magnetar owing to the CME, based on this $\mu_5$, was also studied in Ref.~\cite{SigLei16}. However, analogously to Ref.~\cite{Yam16a}, that model encounters a problem of the small scale of the magnetic field created. 

The system of equations of chiral MHD was formulated in Ref.~\cite{Rog17}; the effects of turbulence in such media were also studied there. The direct numerical simulations of the chiral MHD dynamos were performed in Ref.~\cite{Rog18}. The implication of the obtained results for the description of the magnetic fields generation in proto-NS was discussed in Refs.~\cite{Rog17,Rog18}.

A possible explanation of magnetar bursts~\cite{MerPonMel15} based on the magnetic field reconnection owing to accounting of quantum terms in the magnetic helicity evolution was recently proposed in Ref.~\cite{DvoSem18b}. Note that these quantum corrections to the classical MHD arise from the nonzero mass terms in the Adler-Bell-Jackiw anomaly. It means that one does not need to consider the chiral symmetry restoration in the model in Ref.~\cite{DvoSem18b}.

\section*{Acknowledgments}

I thank the Tomsk State University Competitiveness Improvement Program and RFBR (Grant No.~18-02-00149a) for partial support. 


\begin{thebibliography}{100}

\bibitem{PesSch95}
  M.~E.~Peskin and D.~V.~Schr\"oder,
  \textit{An Introduction to Quantum Field Theory}
  (Reading, Massachusetts, 1995), pp.~659\textendash 667.

\bibitem{NieNin81}
  H.~B.~Nielsen and M.~Ninomiya,
  The Adler-Bell-Jackiw anomaly and Weyl fermions in a crystal,
  Phys. Lett. \textbf{130B}, 389\textendash 396 (1983).

\bibitem{FukKhaWar08}
  K.~Fukushima, D.~E.~Kharzeev, and H.~J.~Warringa,
  The chiral magnetic effect,
  Phys. Rev. D \textbf{78}, 074033 (2008),
  arXiv:0808.3382.

\bibitem{Vil79}
  A.~Vilenkin,
  Macroscopic parity violating effects:
  Neutrino fluxes from rotating black holes and in rotating thermal radiation,
  Phys. Rev. D \textbf{20}, 1807\textendash 1812 (1979).

\bibitem{Sig17}
  G.~Sigl,
  \textit{Astroparticle Physics: Theory and Phenomenology}
  (Atlantis, Paris, 2017).

\bibitem{Koc17}
  V.~Koch, S.~Schlichting, V.~Skokov, P.~Sorensen, J.~Thomas, S.~Voloshin, G.~Wang, and
  H.-U.~Yee,
  Status of the chiral magnetic effect and collisions of isobars,
  Chin. Phys. C \textbf{41}, 072001 (2017),
  arXiv:1608.00982.

\bibitem{BubGubZhu17}
  A.~F.~Bubnov, N.~V.~Gubina, and V.~Ch.~Zhukovsky,
  Vacuum current induced by an axial-vector condensate
  and electron anomalous magnetic moment in a magnetic field,
  Phys. Rev. D \textbf{96}, 016011 (2017).

\bibitem{AkaYam13}
  Y.~Akamatsu and N.~Yamamoto,
  Chiral plasma instabilities,
  Phys. Rev. Lett. \textbf{111}, 052002 (2013),
  arXiv:1302.2125.

\bibitem{BoyRucSha12}
  A.~Boyarsky, O.~Ruchayskiy, and M.~Shaposhnikov,
  Long-range magnetic fields in the ground state of the standard model plasma,
  Phys. Rev. Lett. \textbf{109}, 111602 (2012),
  arXiv:1204.3604.

\bibitem{DvoSem14}
  M.~Dvornikov and V.~B.~Semikoz,
  Instability of magnetic fields in electroweak plasma driven by neutrino asymmetries,
  J. Cosmol. Astropart. Phys. 05 (2014) 002,
  arXiv:1311.5267.

\bibitem{JacKos99}
  R.~Jackiw and V.~A.~Kosteleck\'y,
  Radiatively induced Lorentz and CPT violation in electrodynamics,
  Phys. Rev. Lett. \textbf{82}, 3572\textendash 3575 (1999),
  hep-ph/9901358.


\bibitem{GolSon15}
  S.~Golkar and T.~D.~Son,
  (Non)-renormalization of the chiral vortical effect coefficient,
  J. High Energy Phys.02 (2015) 169,
  arXiv:1207.5806.

\bibitem{BuiPuhVal}
  P.~V.~Buividovich, M.~Puhr, and S.~N.~Valgushev,
  Chiral magnetic conductivity in an interacting lattice model
  of parity-breaking Weyl semimetal,
  Phys. Rev. B \textbf{92}, 205122 (2015),
  arXiv:1505.04582.

\bibitem{Vil80}
  A.~Vilenkin,
  Equilibrium parity-violating current in a magnetic field,
  Phys. Rev. D \textbf{22}, 3080\textendash 3084 (1980).

\bibitem{DvoSem15a}
  M.~Dvornikov and V.~B.~Semikoz,
  Magnetic field instability in a neutron star driven
  by the electroweak electron-nucleon interaction versus the chiral magnetic effect,
  Phys. Rev. D \textbf{91}, 061301 (2015),
  arXiv:1410.6676.

\bibitem{DvoSem15b}
  M.~Dvornikov and V.~B.~Semikoz,
  Generation of the magnetic helicity in a neutron star driven
  by the electroweak electron-nucleon interaction,
  J. Cosmol. Astropart. Phys. 05 (2015) 032,
  arXiv:1503.04162.

\bibitem{KapRedSen17}
  D.~B.~Kaplan, S.~Reddy, and S.~Sen,
  Energy conservation and the chiral magnetic effect,
  Phys. Rev. D \textbf{96}, 016008 (2017),
  arXiv:1612.00032.

\bibitem{DvoSem18}
  M.~Dvornikov and V.~B.~Semikoz,
  Chiral magnetic effect in the presence of electroweak interactions
  as a quasiclassical phenomenon,
  Mod. Phys. Lett. A \textbf{33}, 1850043 (2018),
  arXiv:1702.06426.

\bibitem{SadIsa11}
  A.~V.~Sadofyev and M.~V.~Isachenkov,
  The chiral magnetic effect in hydrodynamical approach,
  Phys. Lett. B \textbf{697}, 404\textendash 406 (2011),
  arXiv:1010.1550.

\bibitem{SonSur09}
  D.~T.~Son and P.~Sur\'owka,
  Hydrodynamics with triangle anomalies,
  Phys. Rev. Lett. \textbf{103}, 191601 (2009),
  arXiv:0906.5044.

\bibitem{BalPopStu11}
  I.~A.~Balantsev, Yu.~V.~Popov, and A.~I.~Studenikin,
  On the problem of relativistic particles motion in strong magnetic field and dense matter,
  J. Phys. A \textbf{44}, 255301 (2011),
  arXiv:1012.5592.

\bibitem{Dvo16a}
  M.~Dvornikov,
  Role of particle masses in the magnetic field generation driven by the parity violating interaction,
  Phys. Lett. B \textbf{760}, 406\textendash 410 (2016),
  arXiv:1608.04940.

\bibitem{ItzZub80}
  C.~Itzykson and J.-B.~Zuber,
  \textit{Quantum Field Theory}
  (McGraw-Hill, New York, 1980),
  pp.~691\textendash 696.

\bibitem{Sil99}
  L.~O.~Silva, R.~Bingham, J.~M.~Dawson, J.~T.~Mendon\c{c}a, and P.~K.~Shukla,
  Neutrino driven streaming instabilities in a dense plasma,
  Phys. Rev. Lett. \textbf{83}, 2703\textendash 2706 (1999).

\bibitem{Lan16}
  K.~Landsteiner,
  Notes on anomaly induced transport,
  Acta Phys. Polon. B \textbf{47}, 2617\textendash 2673 (2016),
  arXiv:1610.04413.

\bibitem{MerPonMel15}
  S.~Mereghetti, J.~Pons, and A.~Melatos. 
  Magnetars: Properties, origin and evolution,
  Space Sci. Rev. \textbf{191}, 315--338 (2015), 
  arXiv:1503.06313.

\bibitem{Yam16a}
  N.~Yamamoto,
  Chiral transport of neutrinos in supernovae:
  Neutrino-induced fluid helicity and helical plasma instability,
  Phys. Rev. D \textbf{93}, 065017 (2016),
  arXiv:1511.00933.

\bibitem{Yam16b}
  N.~Yamamoto,
  Scaling laws in chiral hydrodynamic turbulence,
  Phys. Rev. D \textbf{93}, 125016 (2016),
  arXiv:1603.08864.

\bibitem{SigLei16}
  G.~Sigl and N.~Leite,
  Chiral magnetic effect in protoneutron stars and magnetic field spectral evolution,
  J. Cosmol. Astropart. Phys. 01 (2016) 025,
  arXiv:1507.04983.

\bibitem{Rog17}
  I.~Rogachevskii, O.~Ruchayskiy, A.~Boyarsky,  J.~Fr\"ohlich, N.~Kleeorin,
  A.~Brandenburg, and J.~Schober,
  Laminar and turbulent dynamos in chiral magnetohydrodynamics. I. Theory,
  Astrophys. J. \textbf{846}, 153 (2017),
  arXiv:1705.00378.

\bibitem{Rog18}
  J.~Schober, I.~Rogachevskii, A.~Brandenburg,  A.~Boyarsky, J.~Fr\"ohlich,
  O.~Ruchayskiy, and N.~Kleeorin, 
  Laminar and turbulent dynamos in chiral magnetohydrodynamics. II. Simulations,
  Astrophys. J. \textbf{858}, 124 (2018),
  arXiv:1711.09733.

\bibitem{DvoSem18b}
  M.~Dvornikov and V.~B.~Semikoz,
  Magnetic helicity evolution in a neutron star accounting for the Adler-Bell-Jackiw anomaly,
  J. Cosmol. Astropart. Phys. 08 (2018) 021,
  arXiv:1805.04910.

\end{thebibliography}
\end{document}